\documentclass[aps,prd,twocolumn,floatfix,nofootinbib,superscriptaddress]{revtex4-1}
\usepackage{dcolumn}

\usepackage{amsmath,amsfonts,extarrows,amssymb,mathrsfs,times,bm}
\usepackage{extarrows}
\usepackage{graphicx}
\usepackage{float}
\usepackage{graphicx,epstopdf}
\usepackage{dcolumn}
\usepackage{exscale}
\usepackage{relsize}
\usepackage{mathtools}
\usepackage{xcolor}
\usepackage{mhchem}
\usepackage[colorlinks=true,breaklinks=true,linkcolor=blue,citecolor=blue,urlcolor=blue]{hyperref}
\newcommand{\nn}{\nonumber}

\newcommand{\lb}{\left(}
\newcommand{\rb}{\right)}
\newcommand{\lsb}{\left[}
\newcommand{\rsb}{\right]}

\begin{document}
\title{Deflection of charged signals in a dipole magnetic field in Schwarzschild background using Gauss-Bonnet theorem}
\author{Zonghai Li}
\affiliation{Center for Astrophysics, School of Physics and Technology, Wuhan University, Wuhan 430072, China}

\author{Wei Wang}
\email{wangwei2017@whu.edu.cn}
\affiliation{Center for Astrophysics, School of Physics and Technology, Wuhan University, Wuhan 430072, China}

\author{Junji Jia}
\email{junjijia@whu.edu.cn}
\affiliation{Center for Astrophysics \& MOE Key Laboratory of Artificial Micro- and Nano-structures, School of Physics and Technology, Wuhan University, Wuhan, 430072, China}

\date{\today}

\begin{abstract}
This paper studies the deflection of charged particles in a dipole magnetic field in Schwarzschild spacetime background in the weak field approximation. To calculate the deflection angle, we use Jacobi metric and Gauss-Bonnet theorem. Since the corresponding Jacobi metric is a Finsler metric of Randers type, we use both the osculating Riemannian metric method and  generalized Jacobi metric method. The deflection angle up to fourth order is obtained and the effect of the magnetic field is discussed. It is found that the magnetic dipole will increase (or decrease) the deflection angle of a positively charged signal when its rotation angular momentum is parallel (or antiparallel) to the magnetic field. It is  argued that the difference in the deflection angles of different rotation directions can be viewed as a Finslerian effect of the non-reversibility of the Finsler metric. The similarity of the deflection angle in this case with that for the Kerr spacetime allows us to directly use the gravitational lensing results in the latter case. The dependence of the apparent angles on the magnetic field suggests that by measuring these angles the magnetic dipole might be constrained.
\end{abstract}


\maketitle

\section{Introduction}

The study of particle motions in the gravitational field leads to important discoveries. In particular, trajectory deflection and gravitational lensing (GL) have become important tools in astrophysics. They are used not only in measuring the mass of galaxies and clusters but also in detecting dark matter and dark energy \cite{Book2018}. In addition to the traditional null messengers (i.e. light rays) in GL, in recent years, the deflection and GL of  particles with nonzero mass has also aroused the continuous interest of researchers ~\cite{AR2002,AP2004,Bhadra2007,Yu2014,He&lin1,He&lin2,He&lin3,LZLH2019,Jia2019,JiaKerr}.
Besides the common neutral particles such as photons and neutrinos, charged particles, such as cosmic rays, are also common in our Universe. These charged signals can experience not only the gravitational interaction but also the electromagnetic field existing in/near compact celestial bodies. Investigation of motion of charged particles in both gravitational and electromagnetic fields plays an important role in testing the weak cosmic censorship conjecture~\cite{Wald19741,Sorce&Wald2017}, accretion disks rotating around Kerr Black Holes~\cite{Stuchlik2020} and magnetic Penrose process~\cite{Stuchlik2021,Tursunov2019,Gupta2021}. 

The first motivation of this paper is to study the deflection of charged particles in an electromagnetic field in curved spacetime, which is a generalization of the deflection of light and massive neutral particles in pure gravity. 
More specifically, we will concentrate on the motion of charged particles in a weak magnetic field in curved spacetime backgrounds. The magnetic field is weak in the sense that the correction caused by its energy-momentum tensor to the spacetime metric is negligible compared to the background metric caused by the primary matter distribution.
We study this kind of magnetic/gravitational field because there are no well-known exact solutions admitting a non-trivial and physical magnetic field in gravity. 
Such magnetic fields include the cases of a black hole surrounded by a uniform magnetic field~\cite{Wald19742}, 
and a Schwarzschild spacetime with a dipole magnetic field \cite{Petterson1974}. The latter is particularly interesting because it is believed that neutron stars might allow a magnetic dipole component around it \cite{Wasserman&Shapiro,Sengupta1995,Beskin2018}.
The motion of the charged test particles in this spacetime was then studied in Refs.~\cite{Bakala2010} for the non-geodesic corrections to the particles' orbital and epicyclic frequencies and in Ref. \cite{Preti2004} for the existence and properties of circular orbits.
As far as we know however, the investigation on the particles' deflection and gravitational lensing (GL) in this situation is still lacking. 

In calculating such deflections, a geometric method based on optical geometry and Gauss-Bonnet (GB) theorem introduced by Gibbons and Werner~\cite{GW2008,Werner2012} has become very popular over the years~\cite{ISOA2016,Jusufiworm, Sakalli2017, Arakida2018, Ali:wormhole, Javed2019,Arakida2018, Kumaran2021, QZ2022}.
The optical geometry corresponding to a stationary (and axisymmetric) spacetime is defined by a Finsler metric of Randers type, but the use of Gauss-Bonnet like theorem on Finsler geometry to calculate deflection angles is an open problem. To overcome this difficulty, two formalisms have been developed in the study of light deflection. The first is the {\it osculating Riemannian metric method} introduced by Naz{\i}m in Ref.~\cite{Nazim1936}, and promoted by Werner for calculating the deflection angle of light rays in the equatorial plane in stationary spacetimes~\cite{Werner2012}. In this formalism, the deflection angle formula is the same as in static spacetime. The second is the {\it generalized optical metric method} introduced by Ono, Ishihara, and Asada in Ref.~\cite{OIA2017}. Furthermore, according to optical metrics in special medium~\cite{CG2018,CGJ2019} or Jacobi metric~\cite{LHZ2020,Li:2019qyb,LA2020, LDJ2022}, these methods were extended to the calculation of the deflection of massive particles. In particular, some authors of the current work studied the deflection of charged particles by a Kerr-Newman lens which intrinsically couples the gravitational and electric fields ~\cite{LiJiaPRD2021}.

The second motivation of this work is to extend our previous studies to the case of the deflection of charged particles by a black hole with a dipole magnetic field. In doing so, we will continue to use the geometric method and more importantly, we will test explicitly whether the  osculating Riemannian metric method suggested in Ref. ~\cite{Werner2012} is truly applicable to the case with a magnetic field. We also hope to reveal how the deflection and GL of charged signals are affected by a magnetic dipole, and if possible, to use the former to constrain the strength of such a magnetic field.

The outline of the paper is as follows. In Sec.~\ref{Preliminaries}, we enumerate the preliminaries,  the osculating Riemannian metric method, the generalized Jacobi metric method, GB theorem, and deflection angle formula. In Sec.~\ref{Schwarzschild-dipole}, we introduce the magnetic dipole in a Schwarzschild background and solve the orbit up to the third order in the weak field limit. In Sec.~\ref{osculating-Riemanian} and Sec.~\ref{Generalized Jacobi} respectively, the osculating Riemannian metric method and generalized Jacobi metric method are used to obtain the deflection angle up to the fourth order in this spacetime. These results, especially the effect of the magnetic dipole on the deflection and GL, are analyzed and discussed in Sec.~\ref{Astrophysical-app}. Throughout the paper, we use the natural units $G = c = 1$ and spacetime signature $(-,+,+,+)$.

\section{Preliminaries}\label{Preliminaries}
\subsection{Finsler geometry and non-reversibility of metric}
Finsler geometry is just the Riemannian geometry without the quadratic \cite{Chern1996}. Let $M$ be a $n$-dimensional smooth manifold. It becomes a Finsler manifold $(M,F)$ if we could equip $M$ with a non-negative function $F$ defined on the tangent bundle ${TM}$, satisfying~\cite{Bao-Chern-Shen2002,Shen2001}:

(1) Regularity: $F$ is smooth on $TM\backslash\{0\}$,

(2) Positive 1-homogeneity: $F(x,\xi y)=\xi F(x, y)$ for all $\xi>0$, where $x\in M,y=y^i\partial_i\in T_x{M}$,

(3) Strong convexity: the Hessian matrix of $F^2$
\begin{align}
  (\widetilde g_{ij})=\left(\frac{1}{2}\frac{\partial^2 F^2}{\partial y^i \partial y^j}\right),
 \label{eq:Hessian}
\end{align}
is positive definite, where $\widetilde g_{ij}$ is called the fundamental tensor of $F$.

The line elements in Finsler space $(M,F)$ can be written as
\begin{align}
 ds=F(x^1,...,x^n,dx^1,...,dx^n).
\end{align} 
If the fundamental tensor is only related to $x$, i.e. $\widetilde g_{ij}(x,y)=\widetilde g_{ij}(x)$, then $F$ becomes the Riemannian metric, with the form
\begin{align}
\label{rieman}
 F=\sqrt{\widetilde g_{ij}(x)y^iy^j},
\end{align} 
and the line element
\begin{align}
 ds=\sqrt{\widetilde g_{ij}(x)dx^idx^j}.
\end{align} 
In other words, Riemannian geometry is Finsler geometry with the quadratic.
A special class of non-Riemannian Finsler metrics is the Randers metric of the form
\begin{align}
  F(x,y)=\sqrt{\alpha_{ij}y^iy^j}+\beta_iy^i~,
 \label{eq:rfdef}
\end{align}
where $\alpha_{ij}$ is a Riemannian metric and $\beta_i$ a one-form on $M$, satisfying positivity and convexity
\begin{align}
|\beta|=\sqrt{\alpha^{i j} \beta_{i} \beta_{j}}<1.
\end{align}
This metric was proposed by Randers when he studied the unification of electromagnetism and gravity~\cite{Randers1941}.

The Finsler metric is non-reversible if its reverse $F(x,-y)$ does not equal $F(x,y)$, i.e.,
\begin{align}
 F(x,-y)\neq F(x,y).
\end{align} 
For Riemannian metric \eqref{rieman}, we have
\begin{align}
 F(x,-y)=\sqrt{\widetilde g_{ij}y^iy^j}=F(x,y).
\end{align}
Thus, the Riemannian metric is reversible while this is usually not true for Randers metric
\begin{align}
 F(x,-y)=\sqrt{\alpha_{ij}y^iy^j}-\beta_iy^i\neq  F(x,y).
\end{align}
Now if $y$ is the tangent along a curve, then the reverse metric $F(x,-y)$ can be thought as the metric of the reverse direction along the curve. 

The (non-)reversibility of the metric can affect the arc length of a fixed curve if we measure along different directions. 
Let $c: [a,b]\rightarrow M$ be a curve, $\lambda \in [a,b]$ and $a\leq\lambda_1<\lambda_2\leq b$, the arc length from $c(\lambda_1)$ to $c(\lambda_2)$ along the curve is
\begin{align}
L(c(\lambda_1), c(\lambda_2))=\int_{\lambda_1}^{\lambda_2}F\left(c(\lambda),\frac{dc(\lambda)}{d\lambda}\right)d\lambda.
\end{align}
Likewise, in the opposite direction, the arc length is
\begin{align}
L(c(\lambda_2), c(\lambda_1))=-\int_{\lambda_2}^{\lambda_1}F\left(c(\lambda),-\frac{dc(\lambda)}{d\lambda}\right)d\lambda.
\end{align}
The non-reversibility of metric $F$ means
\begin{align}
L(c(\lambda_1), c(\lambda_2))\neq L(c(\lambda_2), c(\lambda_1)).
\end{align}

Recently, in Refs.~\cite{Masiello2021,Caponio&Masiello2022} the Sagnac effect in General Relativity is considered as a kind of Finslerian effect due to the  non-reversibility of Randers metric.
In this paper, we will show that the difference between the deflection angles for prograde and retrograde motions also stems from the non-reversibility of the Finsler metric. Therefore, the difference in deflection angles can also be considered as a Finslerian effect.

\subsection{Jacobi-Randers metric and the orbital equations}
The line element of a stationary spacetime can be written as
\begin{align}
\label{sspacetime}
d{s}^2={g}_{tt}(x)dt^2+2g_{ti}(x)dtdx^i+{g}_{ij}(x)dx^idx^j.
\end{align}
Assuming that there is an electromagnetic field described by electromagnetic gauge potential $A_\mu$ in spacetime, then the motion of the charged test particle of mass $m$, charge $q$, and energy $E$ is described by the Lorentz equation~\cite{Tursunov}
\begin{align}
\label{lorenzt}
 \frac{d^2 x^{\rho}}{d \tau^2}+\Gamma_{\mu\nu}^{\rho} \frac{dx^\mu}{d\tau} \frac{dx^\nu}{d\tau}=\frac{q}{m}{F}_\mu^\rho\frac{dx^\mu}{d\tau},
\end{align}
where $\Gamma_{\mu\nu}^{\rho}$ is Christoffel symbols of $g_{\mu\nu}$, $\tau$ is the proper time of the test particle, and electromagnetic field tensor ${F}_{\mu\nu}=\nabla_\mu {A}_\nu-\nabla_\nu {A}_\mu$ with $\nabla$ being Levi-Civita connection. Apparently, the motion of charged particles no longer follows geodesics.

As one of the main tools of geometric dynamics, the Jacobi metric has been widely used to study various mechanical problems in the Newtonian framework~\cite{Pin1975, Awrejcewicz}, and has been extended to curved spacetimes~\cite{Gibbons2016, CGGMW2019, Chanda2019}. The trajectories of a given mechanical system of constant total energy, are geodesic within the Jacobi metric, according to Maupertuis's principle. The Jacobi metric for a charged particle moving in spacetime~\eqref{sspacetime} with an electromagnetic field $A_\mu$ is a Finsler metric of Randers type~\cite{Chanda2019}
\begin{align}
\label{fensleranders}
d\rho=F(x,dx)=\sqrt{\alpha_{ij}dx^idx^j}+\beta_idx^i,
\end{align}
with
\begin{align}
\label{fenslerandersA}
\alpha_{ij}&=\frac{\left(E+qA_t\right)^2+m^2{g}_{tt}}{-{g}_{tt}}\left({g}_{ij}-\frac{g_{ti}g_{tj}}{{g}_{tt}}\right),\\
\label{fenslerandersB}
\beta_i&=qA_i-\left(E+qA_t\right)\frac{g_{ti}}{g_{tt}}.
\end{align}
For convenience, we shall call it the Jacobi-Randers metric. Setting $q=0$ in this leads to the Jacobi-Randers metric for a neutral particle, and further setting $m=0$ it becomes the optical metric. In particular, if $\beta_i=0$ (corresponding to $q=0$ and $g_{ti}=0$, or $A_{i}=0$ and $g_{ti}=0$), it becomes Riemannian metric.

For a four-dimensional stationary and axisymmetric spacetime, we can write metric~\eqref{sspacetime} in Boyer-Lindquist coordinates $(t,r,\theta,\phi)$ as
\begin{align}
\label{sspacetime1}
d{s}^2={g}_{tt}(r,\theta)dt^2+2g_{t\phi}(r,\theta)dtd\phi+{g}_{ij}(r,\theta)dx^idx^j,~~~
\end{align}
where $i,~j=1,2,3$ and
\begin{align}
{g}_{ij}dx^idx^j=g_{rr}dr^2+g_{\theta\theta}d\theta^2+g_{\phi\phi}d\phi^2.\nn
\end{align}
The orbital equation of the particle can be derived according to the Jacobi-Randers metric~\eqref{fensleranders}. The Lagrangian of the charged particle in Jacobi metric space is written as
\begin{align}
\mathcal{L}=F(x,\dot{x})&= \sqrt{\alpha_{ij}^{(3)}\dot{x}^i\dot{x}^j}+\beta_i^{(3)}\dot{x}^i,\\
\end{align}
where $\dot x^i\equiv\frac{dx^i}{d\rho}$. Since this paper is only interested in the case of particles moving on the equatorial plane ($\theta=\pi/2$), the $\theta$ dimension can be dropped in this and this Lagrangian becomes
\begin{align}
\label{huangzhongze}
\mathcal{L}=F(x,\dot{x})&=\sqrt{\alpha_{ij}^{(2)}\dot{x}^i\dot{x}^j}+\beta_i^{(2)}\dot{x}^i.
\end{align}
Due to the conservation of angular momentum $J$, we have
\begin{align}
\label{huangjingren}
p_\phi=\frac{\partial \mathcal{L}}{\partial \dot{\phi}}=\beta_\phi^{(2)}+\frac{\alpha_{\phi\phi}^{(2)}\dot{\phi}}{\sqrt{\alpha_{ij}^{(2)}\dot{x}^i\dot{x}^j}}=J.
\end{align}
Using $\mathcal{L}=1$, Eq.~\eqref{huangzhongze} transforms to
\begin{align}
\sqrt{\alpha_{ij}^{(2)}\dot{x}^i\dot{x}^j}=1-\beta_\phi^{(2)} \dot{\phi},
\end{align}
which, after using Eq.~\eqref{huangjingren}, becomes
\begin{align}
\beta_\phi^{(2)}+\frac{\alpha_{\phi\phi}^{(2)}\dot{\phi}}{1-\beta_\phi^{(2)} \dot{\phi}}=J,
\end{align}
or equivalently
\begin{align}
\label{orbit1}
\dot{\phi}=\frac{d\phi}{d\rho}=\frac{\beta_\phi^{(2)}-J}{(\beta_\phi^{(2)})^2-\beta_\phi^{(2)} J-\alpha_{\phi\phi}^{(2)}}.
\end{align}

On the other hand, the equation of motion in the radial direction can also be obtained from Lagrangian~\eqref{huangzhongze} as
\begin{align}
\label{orbit2}
\dot{r}^2=\left(\frac{dr}{d\rho}\right)^2=\frac{1}{{\alpha}_{rr}^{(2)}}\left[\left(1-\beta_\phi^{(2)} \dot{\phi}\right)^2-\alpha_{\phi\phi}^{(2)}\dot{\phi}^2\right].
\end{align}
Introducing the inverse radial coordinate $u\equiv \frac{1}{r}$ and after using Eqs.~\eqref{orbit1}, the above becomes
\begin{align}
\label{orbitequation}
\left(\frac{du}{d\phi}\right)^2=u^4\frac{\alpha_{\phi\phi}^{(2)}\left[\alpha_{\phi\phi}^{(2)}-\left(J-\beta_\phi^{(2)}\right)^2\right]}{\alpha_{rr}^{(2)}\left(J-\beta_\phi^{(2)}\right)^2}.
\end{align}
This is the orbital equation of the particle moving on the equatorial plane.

\subsection{Gauss-Bonnet theorem and deflection angle}
\label{GBLi}
In this subsection we shall derive the deflection angle formula by applying the GB theorem to the lensing geometry. 

Let $D$ be a subset of a compact and oriented two-dimensional surface, with Riemannian metric $\hat{g}_{ij}$ and Euler characteristic number $\chi(D)$. Its boundary $\partial{D}$ is formed by a piecewise smooth curve. The jump angle in the $i$-th vertex of $\partial{D}$ denotes $\varphi_i$, in the positive sense. The GB theorem regarding $D$ states~\cite{GW2008, Carmo1976}
\begin{equation}
\label{GBT}
\iint_{D}{K}d S+\oint_{\partial{D}}{k}_g~dl+\sum_{i}{\varphi_i}=2\pi\chi({D}),\\
\end{equation}
where $K$ is the Gaussian curvature of $D$ and $k_g$ is the geodesic curvature of $\partial{D}$; $dS$ is the area element of $D$ and $dl$ is the line element of $\partial{D}$. Clearly, the GB theorem reveals the relation between the curvature and the topology of $D$.

\begin{figure}[htp!]
\centering
\includegraphics[width=8.0cm]{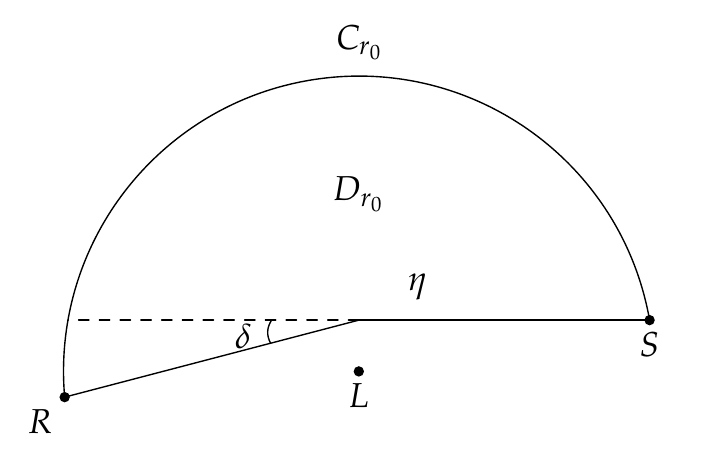}
\caption{The lensing geometry $D_{r_0}$ with boundary $\partial D_{r_0}=\eta \cup C_{r_0}$. $S$, $R$ and $L$ denote the source, the receiver, and the lens, respectively. $\delta$ is the deflection angle. Note that $\varphi_S+\varphi_R\to\pi$ as $r_0\to \infty$ if $M$ is asymptotically Euclidean.\label{Figurelg}}
\end{figure}

Next we will apply the GB theorem to the lensing geometry $D_{r_0}$, a two-dimensional surface that usually can be described by two coordinates ($r,\phi$), as illustrated in Fig. \ref{Figurelg}. Its boundary $\partial D_{r_0}=\eta \cup C_{r_0}$, where $\eta$ is the particle trajectory from the source $S$ to the receiver $R$,  and $C_{r_0}$ is a curve defined by $r=r_0$ with $r_0$ being a large enough constant. Notice that the non-singularity of the region leads to $\chi(D_{r_0})=1$. 

Applying the GB theorem to region $D_{r_0}$, we have
\begin{align}
\label{GBTlibai}
&\iint_{D_{r_{0}}} K d S-\int_{S}^{R}  k_g(\eta) d  l\nn\\
&+\int_{\phi_{S}}^{\phi_{R}} \left.\left[ k_g\frac{d l}{d\phi}\right]\right|_{C_{r_0}}d\phi+\varphi_{R}+\varphi_{S}=2 \pi.
\end{align}
Without losing any generality, we can fix the coordinate system such that $\phi_S=0$ and $\phi_R=(\pi+\delta)$ when $r_0\to \infty$. Here $\delta$ is the small deflection angle that we will attempt to find in this work. The two jump angles clearly satisfy $\varphi_R+\varphi_S\to\pi$ in the limit $r_0\to\infty$. Taking this limit, Eq.~\eqref{GBTlibai} becomes
\begin{align}
\label{shalihuayuan}
&\iint_{D_{\infty}}  K d  S- \lim_{r_0\to \infty}\int_{S}^{R}  k_g(\eta) d  l\nn\\
&+ \int_{0}^{\pi+\delta}\left. \lim_{r_0\to \infty}\left[ k_g\frac{d l}{d\phi}\right]\right|_{C_{r_0}}d \phi=\pi.
\end{align}
To solve the deflection angle, we will have to deal with the integrals in this equation first. 

We first consider the geodesic curvature of the curve $C_{r_0}$. We concentrate on the case that the lensing geometry $D_{r_0}$ is asymptotically Euclidean, i.e., we assume its metric has the following limit when $r\to\infty$
\begin{align}
&d{l}^2\to W^2\left(dr^2+r^2d\phi^2\right),
\end{align}
where $W$ is a constant. We will also assume that the source $S$ and receiver $R$ are in the asymptotically Euclidean region. Therefore, in the limit $r_0\to \infty$, we have
\begin{align}
{k}_g(C_{r_0})\to \frac{1}{Wr_0},~
\left.\frac{d{l}}{d\phi}\right|_{C_{r_0}}\to Wr_0,\end{align}
which yields
\begin{align}
\left.\lim _{r_0 \to \infty}\left( k_{g} \frac{d l}{d \phi}\right)\right|_{C_{r_0}}=1.
\end{align}
Substituting this into Eq.~\eqref{shalihuayuan}, the deflection angle $\delta$ is expressed as
\begin{align}
\label{dashi1}
\delta=-\iint_{D_{\infty}}  K dS+\lim_{r_0\to \infty}\int_{S}^{R}  k_g(\eta) d l.
\end{align}

For the geodesic curvature along $\eta$, then clearly if $\eta$ is a geodesic we would have $ k_g(\eta)=0$ and the deflection angle in this case simplifies to
\begin{align}
\label{dashi2}
\delta=-\iint_{D_{\infty}}  K d S.
\end{align}

Finally, one extra point worth noting is that if the lensing geometry $D_{r_0}$ is not asymptotically Euclidean, one also needs to calculate the geodesic curvature of curve $C_{r_0}$. Choosing its arc length as the parameter, and denoting its tangent as $\dot{C}_{r_0}$, then the geodesic curvature of curve $C_{r_0}$ is~\cite{Jusufiworm}
\begin{align}
k_g( C_{r_0})=\left|\nabla_{\dot{C}_{r_0}} \dot{C}_{r_0}\right|.
\end{align}

In the case of the optical metric or when Jacobi metric is Riemannian, the study of deflection of light or massive particles in the static spacetimes using the GB theorem is straightforward~\cite{GW2008,LHZ2020}. However, if the space in which the particle lives is Finslerian (see Eqs.~\eqref{fensleranders} with \eqref{fenslerandersA} and \eqref{fenslerandersB}), in order to use GB theorem, we need to ``convert'' it to a Riemannian space first. As mentioned in the introduction, there are two formalisms: the osculating Riemannian metric method and the generalized optical metric method. In the next section, we will see that the Jacobi metric corresponding to charged particles in Schwarzschild spacetime with a dipole magnetic field is just a Finsler metric of Randers type. Therefore, in order to study the deflection of charged particles using the GB theorem in this case, the two methods mentioned above will be used in Sec
.~\ref{osculating-Riemanian} and Sec.~\ref{Generalized Jacobi}, respectively.

\section{Schwarzschild spacetime in a dipole magnetic field}\label{Schwarzschild-dipole}

In this section, we introduce the Schwarzschild spacetime in a dipole magnetic field. This configuration is useful in modeling many spherical mass distributions with dipole magnetic field,  including pulsars (or even planets) whose magnetic field is generated by internal matter/charge flows, and compact objects (including neutron stars and BHs) with an accretion that might contribute to the magnetic field.

\subsection{Schwarzschild metric with magnetic field \& induced Jacobi-Randers metric}

In flat spacetime, assuming the magnetic dipole is centered at the origin and oriented along the $z$ direction, then its vector form in $(r,\theta,\phi)$ coordinates can be written as
\begin{align}
\mathbf{B}=\frac{2 \mu}{r^{3}}\left(\cos \theta, \frac{1}{2} \sin \theta, 0\right).
\label{eq:flatmag}
\end{align}
where $\mu$ is the magnetic dipole moment.

In Schwarzschild spacetime described by the line element
\begin{align}
\label{spacetime.geometry}
& d{s}^2=-f(r)dt^2+\frac{dr^2}{f(r)}+r^2\left(d\theta^2+\sin^2\theta d\phi^2\right),
\end{align}
where 
\begin{align}
f(r)\equiv 1-\frac{2M}{r}
\end{align}
however, the dipole magnetic field needs to be modified if it is generated by a loop of current $I$ and of radius $R_0$ on the equatorial plane~\cite{Petterson1974}.
The generating electromagnetic potential $A_\mu$ has only one nonzero component $A_\phi$, which takes the form
\begin{align}
& A_\phi=-\frac{3}{8}\frac{\mu r^2\sin^2\theta}{M^3}\left[\ln f(R_0)+\frac{2M}{R_0}\left(1+\frac{M}{R_0}\right)\right]
\end{align}
in the inner region of $2M< r< R_0$ and
\begin{align}
\label{electr.potential}
 &A_\phi=-\frac{3}{8}\frac{\mu r^2\sin^2\theta}{M^3}\left[\ln f(r)+\frac{2M}{r}\left(1+\frac{M}{r}\right)\right],
\end{align}
in the outer region of $r\geq R_0$. Here 
$ \mu=\pi I R_0^2\sqrt{f(R_0)}$
is the asymptotic dipole moment of the field. 
In this work, since we only concentrate on the weak field limit, it is the external vector potential \eqref{electr.potential} that will be used. 
On the equatorial plane, the potential \eqref{electr.potential} generates a magnetic field in the local Lorentz frame with the only nonzero component 
\begin{align}
    B_\theta=3\mu 
    \frac{\sqrt{f(r)}}{4M^3}\left[\ln f(r)+\frac{2M}{r}\frac{\left(1-M/r\right)}{f(r)}\right],
    \label{eq:btheta}
\end{align}
When $(r\gg R_0>2M)$, this magnetic field becomes its asymptotic value given in Eq. \eqref{eq:flatmag}. Note that in this work, we do not restrict the sign of $\mu$, i.e., $\mu$ can be negative so that the dipole is pointed to the $-\hat{z}$ direction. 

Using spacetime geometry~\eqref{spacetime.geometry} and electromagnetic potential~\eqref{electr.potential}, we can derive the corresponding Jacobi metric for a charged test particle. Noting the facts that $g_{t\phi}=0$ in the metric and $A_t=A_r=A_\theta=0$ in the electromagnetic potential, Jacobi metric Eq.~\eqref{fensleranders} is found to be
\begin{align}
\label{frdata}
d\rho=F(x,dx)=\sqrt{\alpha_{ij}^{(3)}dx^idx^j}+\beta_i^{(3)} dx^i,
\end{align}
with 
\begin{subequations}
\label{eq:alphabeta3}
\begin{align}
\label{FRdata1}
\alpha_{ij}^{(3)}&dx^idx^j=\left( E^2 f(r)^{-1}-m^2\right)\nn\\
&\times\bigg[f(r)^{-1}dr^2+r^2\left(d\theta^2+\sin^2\theta d\phi^2\right)\bigg],\\
\label{FRdata2}
\beta_i^{(3)}&dx^i=-\frac{3}{8}\frac{q\mu r^2\sin^2\theta}{M^3}\nn\\
&\times\left[\ln f(r)+\frac{2M}{r}\left(1+\frac{M}{r}\right)\right] d\phi.
\end{align}
\end{subequations}
Due to the presence of the nonzero magnetic dipole moment $\mu$, this metric is a Finsler metric of Randers type, which is different from previous observations, where the Jacobi metric for neutral or charged particles in the static spacetime is merely a Riemannian metric~\cite{LHZ2020, LDJ2022}.

Concentrating on the equatorial plane  ($\theta=\frac{\pi}{2},d\theta=0$), the above Jacobi-Randers metric \eqref{frdata} with \eqref{FRdata1} and \eqref{FRdata2} becomes
\begin{align}
\label{frdata1}
d\rho=&F(r,\phi, dr,d\phi)\nn\\
=&\sqrt{\alpha_{ij}^{(2)}dx^idx^j}+\beta_\phi^{(2)}dx^i\nn\\
=&\sqrt{\alpha_{rr}^{(2)}dr^2+\alpha_{\phi\phi}^{(2)}d\phi^2}+\beta_\phi^{(2)} d\phi,
\end{align}
with 
\begin{subequations}
\label{eq:metricalpha2beta}
\begin{align}
\label{frdata11}
\alpha_{rr}^{(2)}=&\left[E^2-m^2f(r)\right]f(r)^{-2},\\
\label{frdata12}
\alpha_{\phi\phi}^{(2)}=&\alpha_{rr}r^2f(r),\\
\label{frdata13}
\beta_\phi^{(2)}=&-\frac{3}{8}\frac{q\mu r^2}{M^3}\left[\ln f(r)+\frac{2M}{r}\left(1+\frac{M}{r}\right)\right].
\end{align}
\end{subequations}
Next we will establish the orbital equations using this metric. 
If we substitute Eqs. \eqref{eq:alphabeta3} or \eqref{eq:metricalpha2beta} into the equations of motion \eqref{orbit1} and \eqref{orbit2}, then observing that $\beta_i^{(2)}$ is linear to $\mu$ while $\alpha_{ij}^{(2)}$ is independent of $\mu$, it is clear that switching the particle's motion direction from anti-clockwise to clockwise, i.e., $s=+1\to s=-1$ and $J>0\to J^\prime =-J$ (see Eq. \eqref{eq:binj}), will be equivalent to keeping $s$ while change the sign of $\mu$. This implies that the deflection angle for fixing $\mu$ but switching $s$ would also be the same as keeping $s$ and changing the sign of $\mu$. Without losing any generality, in the following, we will concentrate on the case $s=+1$ and restore the sign to $\mu$ in the final expression.

From the geometrical point of view. for the motion in the reverse direction, the deflection angle then can be studied using the reverse metric $F(x,-dx)$. Inspecting $F(x,dx)$ and $ F(x,-dx)$, we find that we can obtain the deflection angle in the reverse direction by simply replacing $\mu$ with $-\mu$ (or $q$ with $-q$) in the deflection angle along the prograde direction. This indicates that the difference in deflection angle originates from the non-reversibility of the Finsler metric. In other words, this difference is a Finslerian effect.

\subsection{Orbital equations of charged particle}\label{litaibai}

Substituting the metric functions in Eq.~\eqref{frdata1} into Eq.~\eqref{orbitequation}, the orbital equation for a charged particle moving on the equatorial plane in Schwarzschild spacetime with a dipole magnetic field can be obtained, as follows
\begin{align} \label{eq:uinbetaphi}
\left(\frac{du}{d\phi}\right)^2=-\left(1-2Mu\right)u^2+\frac{E^2-m^2\left(1-2Mu\right)}{\left(\beta_\phi^{(2)}-J\right)^2}.~~~~
\end{align}
Since we are interested in the weak field limit, this equation can be solved perturbatively according to the asymptotic condition $\lim_{\phi \to 0} u=0$~\cite{CGJ2019}; that is, the particle asymptotically approaches the $\phi=0$ radial direction. 
Using the undetermined coefficient method, the solution to Eq. \eqref{eq:uinbetaphi} up to the third order of $(M/b)$ or $(q\mu/b^2)$ is found to be 
\begin{align}
\label{orbit3}
u_M=&u_0+u_1\frac{M}{b}+u_2\frac{M^2}{b^2}+u_3\frac{M^3}{b^3}\nn\\
&+u_4\frac{q\mu}{b^2}+u_5 \frac{Mq\mu}{b^3}+\mathcal{O}\left(\frac{[M]^4}{b^4}\right),
\end{align}
where coefficients $u_i$ are
\begin{subequations}
\label{eq:orbitsols}
\begin{align}
\label{orbit30}
u_0=&\frac{\sin\phi}{b},\\
\label{orbit31}
u_1=&\frac{(1-\cos\phi)\left(1-v^{2} \cos\phi\right)}{b v^{2}},\\
\label{orbit32}
u_2=&-\frac{1}{8 b v^{2}} \cos\phi\bigg\{6\left(4+v^{2}\right) \phi-16\left(1+v^{2}\right) \sin\phi\nn\\
&+\left[-8+7 v^{2}+3 v^{2} \cos(2\phi)\right] \tan\phi\bigg\},\\
\label{orbit33}
u_3=&-\frac{1}{8 b v^{4}}\left(1+v^{2}-2 v^{2} \cos\phi\right) \sin\left(\frac{\phi}{2}\right)\nn\\
&\bigg\{12\left(4+v^{2}\right) \phi \cos\left(\frac{\phi}{2}\right)-2\sin\left(\frac{\phi}{2}\right)\nn\\
&\left[40+7 v^{2}+\left(8+6 v^{2}\right) \cos\phi-v^{2} \cos(2\phi)\right] \bigg\},\\
\label{orbit34}
u_4=&\frac{\sqrt{1-v^{2}}(1-\cos\phi)}{b m v},\\
\label{orbit35}
u_5=&-\frac{\sqrt{1-v^{2}}}{2 b m v^{3}}\bigg[\left(4+5 v^{2}\right) \phi \cos\phi\nn\\
&-\left(4+v^{2}+4 v^{2} \cos\phi\right) \sin\phi\bigg].
\end{align}
\end{subequations}
Here $b$ is the impact parameter satisfying 
\begin{align}
b=\frac{J}{sEv}, \label{eq:binj}
\end{align}
and we have used
\begin{align}
E=\frac{m}{\sqrt{1-v^2}}, \quad J=\frac{sbvm}{\sqrt{1-v^2}}, 
\label{eq:ejinvb}
\end{align}
with $v$ the asymptotic velocity of the particle. 

\section{Deflection angle using the osculating Riemannian metric method}\label{osculating-Riemanian}

The osculating Riemannian metric method developed by Werner has proven successful in the studies of the deflection of light or particles in stationary spacetime~\cite{Werner2012,CGJ2019,LiJiaPRD2021}. This paper deals with a completely new situation: the spacetime background is static, and the charged particles are simultaneously affected by the gravitational field (Schwarzschild spacetime) and the external magnetic field (a dipole magnetic field). Our purpose in this section is to show that the method is also useful in this case. 

The advantage of Werner's method is that the particle's trajectory is geodesic in the osculating Riemannian manifold, thus we do not need to consider the effect of geodesic curvature on deflection angle. This allows us to use Eq.~\eqref{dashi2} to calculate the deflection angle if the osculating Riemannian space is asymptotically Euclidean. The disadvantage of this method is that the calculation is cumbersome, thus we only consider the deflection angle up to the second order of $(M/b)$ or $(q\mu/b^2)$.

\subsection{Osculating Riemannian metric method}
The fundamental tensor of a Finsler metric is the Hessian of ${F}^2$ defined by (also see Eq.~\eqref{eq:Hessian})~\cite{Bao-Chern-Shen2002}
\begin{align}
\label{Hessian}
 \widetilde{g}_{ij}(x,y)=\frac{1}{2}\frac{\partial^2{F}^2(x,y)}{\partial y^i \partial y^j}~.
\end{align}
It is not difficult to verify that the fundamental tensor for Randers metric $(\alpha_{ij},\beta_i)$ can be written as
\begin{align}
\label{baishidaoren}
 \widetilde{g}_{i j}\left(x, y\right)=&\alpha_{i j}+\beta_{i} \beta_{j}+\frac{\left(\alpha_{i j} \beta_{k}+\alpha_{j k} \beta_{i}+\alpha_{k i} \beta_{j}\right) y^{k}}{\left(\alpha_{k l} y^{k} y^{l}\right)^{1 / 2}}\nn\\
&-\frac{\left(\beta_{k} y^{k}\right) \alpha_{i k} \alpha_{j l} y^{k} y^{l}}{\left(\alpha_{k l} y^{k} y^{l}\right)^{3 / 2}}.
\end{align}

The idea is that we can choose a smooth nonzero vector field $Y(x)$ over ${M}$, such that its restriction on the geodesic $\eta_{{F}}$ is exactly the tangent vector $\dot{\eta}_{{F}}$, i.e., $Y(\eta_F)=\dot{\eta}_{{F}}$. Then the osculating Riemannian metric along the geodesic is defined by
\begin{align}
\label{lingling}
 \bar{g}_{ij}(x)= \widetilde{g}_{ij}\left(x,Y(x)\right)~.
\end{align}
In this construction, the geodesic in $({M}, {F})$ is also a geodesic in $({M}, \bar{g})$~\cite{Werner2012}. Let $D_{r_0}\subset(M, \bar{g})$ (See Ref.~\cite{Werner2012} for details), one then can use the GB theorem to study the deflection of particles (see the derivation in subsection~\ref{GBLi}).

\subsection{Schwarzschld-Dipole Jacobi-Randers Osculating Riemannian metric}
First substituting $y^i$ for $dx^i$ in Eq. \eqref{frdata1}, the Finsler-Randers metric of the Schwarzschild spacetime with a magnetic dipole field becomes
\begin{align}
&F(r,\phi,y^r,y^\phi)=\sqrt{\alpha_{ij}^{(2)}y^iy^j}+\beta_i^{(2)}y^i\nn\\
=&\sqrt{\alpha_{rr}^{(2)}\left(y^r\right)^2+\alpha_{\phi\phi}^{(2)}\left(y^\phi\right)^2}+\beta_\phi^{(2)} y^\phi,
\label{eq:ffwithy}
\end{align}
where $\alpha_{rr}^{(2)}$, $\alpha_{\phi \phi}^{(2)}$ and $\beta_\phi^{(2)}$ are given by Eqs.~\eqref{eq:metricalpha2beta}.
Further substituting this into Eq.~\eqref{baishidaoren} and then using Eq.~\eqref{lingling}, the osculating Riemannian metric of Randers metric $(\alpha_{ij}^{(2)},\beta_i^{(2)})$ becomes
\begin{align}
\label{osc-Riem}
 \bar{g}_{ij}(r,\phi)=& \widetilde{g}_{ij}\left(r,\phi,Y^r(r,\phi),Y^\phi(r,\phi)\right)\nn\\
 =&\alpha_{i j}^{(2)}+\beta_{i}^{(2)} \beta_{j}^{(2)}
 -\frac{\left(\beta_{m}^{(2)} Y^{m}\right) \alpha_{i n}^{(2)} \alpha_{j p}^{(2)} Y^{n} Y^{p}}{\left(\alpha_{k l}^{(2)} Y^{k} Y^{l}\right)^{3 / 2}}\nn\\
 +&\frac{\left(\alpha_{i j}^{(2)} \beta_{k}^{(2)}+\alpha_{j k}^{(2)} \beta_{i}^{(2)}+\alpha_{k i}^{(2)} \beta_{j}^{(2)}\right) Y^{k}}{\left(\alpha_{k l}^{(2)} Y^{k} Y^{l}\right)^{1 / 2}},
\end{align}
where we choose $Y\equiv (\dot{r},\dot{\phi})$, the vector field along the geodesic.

In order to compute this $\bar{g}_{ij}$ to the order of $(M/r)^2$ or $(q\mu/r^2)$, we first note that the $\beta_i^{(2)}$ given in Eq.~\eqref{frdata13} is already at the order $(M/r)^2$ or $(q\mu/r^2)$,
\begin{align}
\beta_\phi^{(2)}=\frac{\mu}{r}+\frac{3M \mu}{2r^2}+\frac{12M^2\mu}{5r^3}+\mathcal{O}\left(\frac{[M]^5}{r^4}\right).
\end{align}
Thus, for the tangent field $Y$ we only need to compute it using the lowest order geodesics, i.e., Eq. \eqref{orbit30} or equivalently $r=b/\sin \phi$. Then, using Eqs.~\eqref{orbit1} and~\eqref{orbit2}, $Y$ becomes
\begin{align}
\label{vectorfield}
Y=(\dot{r},\dot{\phi})=\left(-\frac{\cos \phi}{E v},\frac{ \sin ^{2} \phi}{Eb v}\right).
\end{align}
Substituting into Eq.~\eqref{osc-Riem},  the components of osculating Riemannian metric up to order $(M/r)^2$ or $(q\mu/r^2)$ are found to be
\begin{subequations}
\label{eq:bargij}
\begin{align}
\bar g_{rr} =&E^2v^2+2\left(1+{v}^{2}\right)\frac{E^2M}{r}+4\left(2+v^2\right)\frac{E^2M^2}{r^2}\nn\\
&+\frac{Evrq\mu}{b^3}\frac{ \sin^6\phi}{\chi^{\frac{3}{2}}}+\mathcal{O}\left(\frac{[M]^3}{r^3}\right),\\
\bar{g}_{r\phi}=&\bar{g}_{\phi r}=-\frac{Ev q\mu}{r}\frac{\cos^3\phi}{\chi^{\frac{3}{2}}}+\mathcal{O}\left(\frac{[M]^3}{r^2}\right),\\
\bar{g}_{\phi\phi}=&E^2r^2v^2+2E^2Mr+4E^2M^2\nn\\
&+\frac{Evrq\mu}{b}\frac{\left(\cos^2\phi+2\chi\right)\sin^2\phi}{\chi^{\frac{3}{2}}}+\mathcal{O}\left(\frac{[M]^3}{r^3}\right),
\end{align}
\end{subequations}
where $\chi=\cos^2 \phi+r^2 \sin^4 \phi/b^2$.

\subsection{The deflection angle}
In the limit of $r\to \infty$, it is straightforward to check that  $\bar{g}_{ij}$ behaves to the leading order of $r$ as
\begin{align}
&\bar{g}_{ij}dx^idx^j\to E^2v^2\left(dr^2+r^2d\phi^2\right).
\end{align}
This implies the osculating Riemannian metric is asymptotically Euclidean and thus we can use Eq.~\eqref{dashi2} to calculate the deflection angle. The Gaussian curvature of Riemannian metric $\bar{g}_{ij}$ is known as~\cite{Werner2012}
\begin{align}
\label{Gausscurvature}
\bar{K}=\frac{1}{\sqrt{\bar g}}\bigg[\frac{\partial}{\partial \phi}\left(\frac{\sqrt{\bar g}}{\bar g_{r r}} \bar \Gamma_{r r}^{\phi}\right)-\frac{\partial}{\partial r}\left(\frac{\sqrt{\bar g}}{\bar g_{r r}} \bar\Gamma_{r \phi}^{\phi}\right)\bigg],
\end{align}
where $\bar g$ and $\bar\Gamma_{ij}^k $ are the determinant and Christoffel symbols of metric $\bar g_{ij}$, respectively.

Substituting Eq. \eqref{eq:bargij} into Eq.~\eqref{Gausscurvature}, the  Gauss curvature of $\bar g_{ij}$ is calculated as
\begin{align}\label{eq:barckres}
&\bar{K} d S=\bar{K}\sqrt{\bar g}d ud\phi\nn\\ =&\left[\left(1+\frac{1}{v^{2}}\right) M +\left(1+\frac{6}{v^{2}}-\frac{4}{v^{4}}\right) M^{2} u\right.\nn\\
&+\frac{3H(1/u,\phi)}{2Ev}\frac{q\mu}{ub^2}+\mathcal{O}\left([M]^3u^2\right)\bigg]dud\phi
\end{align}
where for later easier integration we have changed the integration variable from $r$ to $u=1/r$ and
\begin{align}
H(r, \phi)=&\frac{\sin^3\phi}{\left(\cos^2\phi+\frac{r^{2}}{b^{2}} \sin^4\phi\right)^{\frac{7}{2}}} \bigg[2 \cos^6 \phi\left(\frac{5r}{b} \sin\phi-2\right)\nn\\
&-\cos^4 \phi \sin^2 \phi\left(2-9 \frac{r}{b} \sin \phi+10 \frac{r^{3}}{b^{3}} \sin^3 \phi\right)\nn\\
&+4 \frac{r}{b} \cos^2\phi\sin^5 \phi\left(1+2 \frac{r}{b} \sin\phi-\frac{r^{2}}{b^{2}}\sin^2\phi\right)\nn\\
&+\frac{r^{2}}{b^{2}}\left(-\frac{r}{b} \sin^9\phi+2 \frac{r^{3}}{b^{3}} \sin^{11} \phi+\sin^4(2 \phi)\right)\bigg].\nn
\end{align}
Moreover, for the integral limits of $u$ in Eq.
~\eqref{dashi2}, we can simply use the 
signal trajectory to first order, i.e., $u_M=u_0+u_1M/b$ where $u_0$ and $u_1$ are given in  Eqs.~\eqref{orbit30} and~\eqref{orbit31}, as the upper limit, 
and zero (corresponding to $r\to\infty$) as the lower limit. 

Finally, substituting $\bar {K}$ into Eq.~\eqref{dashi2} and carrying out the double integral, the deflection angle to the order $(M/b)^2$ or $q\mu/b^2$ can be obtained, order by order as in Eq. \eqref{eq:barckres}, as the following
\begin{align}
\label{li2}
\delta=&\int_0^\pi\int_0^{u_0+u_1M/b}\bar{K}dud\phi\nn\\
=&\frac{2 M}{b}\left(1+\frac{1}{v^{2}}\right)+\left(1+\frac{4}{v^{2}}\right) \frac{3\pi M^{2}}{4b^{2}}\nn\\
&+ \frac{2sq \mu}{Evb^{2}}+\mathcal{O}\left(\frac{[M]^3}{b^3}\right).
\end{align}
Here we have restored the sign $s=\pm 1$ for different motion directions. In this way, we show that it is also possible to calculate the deflection angle of charged massive
particles in the magnetic field in curved spacetime using Werner’s Osculating Riemannian metric.

\section{Deflection angle using the generalized Jacobi metric method}\label{Generalized Jacobi}

In the generalized optical/Jacobi metric method, the particles' trajectory is no longer a geodesic, and we need to consider the contribution of the geodesic curvature to the deflection angle. In particular, if the generalized optical/Jacobi metric space is asymptotically Euclidean, we can use Eq.~\eqref{dashi1} to calculate the deflection angle. Although the geodesic curvature term is added, the calculation of this method is not tedious, and it is suitable for the calculation of high-order deflection angle~\cite{CGJ2019}. In order to fully study the effect of magnetic charge on the deflection of charged particles, we shall calculate the deflection angle to the fourth order.

\subsection{Generalized Jacobi metric method}

The motion of a particle in Finsler-Randers space $(\alpha_{ij},\beta_i)$ can be equivalent to the motion of the particle in Riemannian space defined by $\alpha_{ij}$, plus a disturbance of one-form $\beta_i$. Because we work with the Jacobi metric rather than optical metric, we will refer to the method outlined in this part as the {\it generalized Jacobi metric method} in this paper.

This method assumes that the test particles live in the 3-dimensional Riemannian space defined by
\begin{align}
 dl^2=\alpha_{ij}^{(3)}dx^idx^j,
\end{align}
where the Riemannian metric $\alpha_{ij}^{(3)}$ is also called the generalized Jacobi metric.
The arc length $l$ here is an affine parameter along the particle trajectory $\eta$.
The equation of motion in this space then can be written as~\cite{OIA2017, Asada&Kasai2000}
\begin{align}
 \frac{d^2 x^{i}}{d l^2}+{ }^{(3)} \Gamma_{j k}^{i} \frac{dx^j}{dl} \frac{dx^k}{dl}=\alpha^{(3)ij}\left(\nabla_j\beta_{k}^{(3)}-\nabla_i\beta_{j}^{(3)}\right)\frac{dx^k}{dl},
\end{align}
with ${ }^{(3)} \Gamma_{j k}^{i} $ being the 3-dimensional Christoffel symbol associated with $\alpha_{ij}^{(3)}$. Note that
since the right side of the above equation is nonzero, the trajectory $\eta$ is no longer a geodesic in the generalized Jacobi metric space. Indeed, we can rewrite the above equation as
\begin{align}
 \frac{d^2 x^{i}}{d l^2}+{ }^{(3)} \Gamma_{j k}^{i} \frac{dx^j}{dl} \frac{dx^k}{dl}=\mathcal{B}_k^i\frac{dx^k}{dl},
\end{align}
where $\mathcal{B}_{ij}=\nabla_i \mathcal{\beta}_j^{(3)}-\nabla_j \mathcal{\beta}_i^{(3)}$. This equation is similar to the Lorentz equation~\eqref{lorenzt}.
Due to this non-geodesicity, when calculating the deflection angle, we will have to take into account the contribution of geodesic curvature $k_g$. 

\subsection{Gauss curvature and geodesic curvature}
On the equatorial plane, the generalized Jacobi metric is
\begin{align}
\label{tianshantonglao}
dl^2=&\alpha_{ij}^{(2)}dx^idx^j,
\end{align}
where $\alpha_{ij}^{(2)}$ are given by Eqs. \eqref{frdata11} and \eqref{frdata12}. This metric is asymptotically Euclidean, because 
\begin{align}
&\alpha_{ij}^{(2)}dx^idx^j\to E^2v^2\left(dr^2+r^2d\phi^2\right),
\end{align}
in the limit of $r\to \infty$. Therefore, one can use the GB theorem formula ~\eqref{dashi1} to compute the deflection angle.

For the Gaussian curvature term, a direct computation using metric $\alpha_{ij}^{(2)}$ yields
\begin{align}
\label{kdsdata}
K d S =&\bigg[\left(1+\frac{1}{v^{2}}\right) M+\left(1+\frac{6}{v^{2}}-\frac{4}{v^{4}}\right) M^{2} u\nn\\
&+ \frac{3}{2}\left(1+\frac{15}{v^{2}}-\frac{20}{v^{4}}+\frac{8}{v^{6}}\right) M^{3} u^{2}\nn\\
&+\frac{1}{2}\left(5+\frac{140}{v^{2}}-\frac{280}{v^{4}}+\frac{224}{v^{6}}-\frac{64}{v^{8}}\right) M^{4} u^{3}\nn\\
&+\mathcal{O}\left(u^4\right)\bigg]dud\phi,
\end{align}
where again we have changed the integration variable from $r$ to $u=1/r$. For a particle moving in the equatorial plane, the geodesic curvature of the particle ray can be calculated by the following equation~\cite{OIA2017}
\begin{align}
\label{qiaofeng}
k_g(\eta)=\left.\left[-\frac{1}{\sqrt{\alpha^{(3)} {\alpha}^{(3)\theta\theta}}}\frac{\partial }{\partial r}\beta_\phi^{(3)}\right]\right|_{\theta=\pi/2},
\end{align}
where $\alpha^{(3)}=\det (\alpha_{ij}^{(3)})$. A direct calculation using the data of $(\alpha_{ij}^{(3)},\beta_i^{(3)})$ given by Eqs.~\eqref{FRdata1} and~\eqref{FRdata2}, yields
\begin{align}
k_g(\eta)=&\frac{q \mu}{E^{2} v^{2}} u^{3}+\left(1-\frac{1}{v^{2}}\right) \frac{2 M q \mu}{E^{2} v^{2}} u^{4}\nn\\
&+\left(\frac{37}{10}+\frac{4}{v^{4}}-\frac{8}{v^{2}}\right) \frac{M^{2} q \mu}{E^{2} v^{2}} u^{5}+\mathcal{O}\left([M]^5u^6\right).
\end{align}
To integrate this, we will change the integration variable from $l$ to $\phi$ using relation $dl/d\phi$, which can be worked out from line element~\eqref{tianshantonglao} and trajectory solution~\eqref{eq:orbitsols}. After this, we obtain
\begin{align}
\label{yangbailao}
&\kappa_g(\eta)d l=\left[k_g(\eta)\frac{dl}{d\phi}\right] d\phi\nn\\
=
& \left\{\frac{\sin \phi}{Ev} \frac{q \mu}{b^{2}}+\frac{2\left(1+2 v^{2}+v^{2} \cos\phi\right) \sin\phi}{E v^3} \frac{M q \mu}{b^{3}}\right.\nn\\
&+\frac{2 }{ E^{2} v^{2}} (2+\cos \phi)\sin^2\left(\frac{\phi}{2}\right)\frac{q^{2} \mu^{2}}{b^{4} }\nn\\
&+\frac{\sin\phi}{80  E v^{3}}
\bigg[400+238 v^{2}-160\left(1+v^{2}\right) \cos\phi\nn\\
&-18 v^{2} \cos2\phi-60\left(4+v^{2}\right) \phi \cot\phi\bigg]\frac{M^{2} q \mu}{b^{4}}\nn\\
&\left.+\mathcal{O}\left(\frac{[M]^5}{b^5}\right)\right\}d\phi.
\end{align}

\subsection{The deflection angle}
If we substitute Eqs. \eqref{kdsdata} and \eqref{yangbailao} into \eqref{dashi1} to calculate the deflection, when carrying out the integral over $\phi$ however, its upper limit depends on the deflection itself. Therefore to be self-consistent, what we will do is to compute the deflection in the weak field limit iteratively. To do this, it is important to note that when using the GB theorem to calculate the deflection angle, to obtain the $n$-th order deflection angle, we need the $(n-1)$-th order particle orbit and the $(n-2)$-th order deflection angle~\cite{CGJ2019}. Therefore We will first calculate the second order deflection angle using first order orbit and zeroth order deflection, and then increase the order one by one. 
Using the first order orbit $u_M=u_0+u_1M/b$ given by Eqs.~\eqref{orbit30} and \eqref{orbit31} and the zeroth order deflection angle $\delta^{(0)}=0$ in Eq.~\eqref{dashi1}, and substituting  Eqs.~\eqref{kdsdata} and~\eqref{yangbailao} up to second order, 
the deflection angle to the second order then becomes 
\begin{align}
\label{wangchongyang}
\delta=&\int_0^{\pi+\delta^{(0)}}\int_0^{u_0+u_1M/b}\left[ \left(1+\frac{1}{v^{2}}\right) M\right.\nn\\
&\left.+\left(1+\frac{6}{v^{2}}-\frac{4}{v^{4}}\right) M^{2} u\right] dud\phi\nn\\
&+\int_0^{\pi+\delta^{(0)}}\frac{\sin \phi}{Ev} \frac{q \mu}{b^{2}} d\phi\nn\\
=&2\left(1+\frac{1}{v^{2}}\right) \frac{M}{b}+\left[\frac{3 \pi}{4}\left(1+\frac{4}{v^{2}}\right) \frac{M^{2}}{b^{2}}+\frac{2}{E v} \frac{q \mu}{b^{2}}\right]\nn\\
&+\mathcal{O}\left(\frac{[M]^3}{b^3}\right),
\end{align}
where the first and second terms are respectively the first and second order result of the deflection angle. This is consistent with Eq.~\eqref{li2}, the result obtained by the osculating Riemannian metric method. 

With this second order deflection, we can then compute the fourth order deflection angle with the help of the third order orbit given by Eq.~\eqref{orbit3}. Using Eq. \eqref{dashi1}, the Gaussian deflection to the fourth order should be obtained by the full result in Eq. \eqref{kdsdata}
\begin{align}
\delta^{K}=&\int_0^{\pi+\delta^{(1)}+\delta^{(2)}}\int_0^{u_M} K \sqrt{\alpha^{(2)}}dud\phi\nn\\
=&2\left(1+\frac{1}{v^{2}}\right) \frac{M}{b}+\frac{3 \pi}{4}\left(1+\frac{4}{v^{2}}\right) \frac{M^{2}}{b^{2}}\nn\\
&+\frac{2}{3}\left(5+\frac{45}{v^{2}}+\frac{15}{v^{4}}-\frac{1}{v^{6}}\right) \frac{M^{3}}{b^{3}}\nn\\
&+\frac{\pi \sqrt{1-v^{2}}}{m v}\left(1+\frac{1}{v^{2}}\right) \frac{q \mu M}{b^{3}}\nn\\
&+\frac{105 \pi}{4}\left(\frac{1}{16}+\frac{1}{v^{2}}+\frac{1}{v^{4}}\right) \frac{M^{4}}{b^{4}}\nn\\
&+\frac{2 \sqrt{1-v^{2}}}{m v}\left(6+\frac{17}{v^{2}}+\frac{2}{v^{4}}\right) \frac{q \mu M^{2}}{b^{4}}+\mathcal{O}\left(\frac{[M]^5}{b^5}\right),\nn
\end{align}
where $u_M$ is third order obit given by Eq.~\eqref{orbit3} with Eqs.~\eqref{orbit30} -\eqref{orbit35} and $\delta^{(1)}+\delta^{(2)}$ is given by Eq.~\eqref{wangchongyang}.
The deflection due to geodesic curvature to the fourth order is found using the full Eq. \eqref{yangbailao} 
\begin{align}
 \delta^{k_g}=&\int_0^{\pi+\delta^{(1)}+\delta^{(2)}}\left[k_g(\eta)\frac{dl}{d\phi}\right] d\phi\nn\\
 =&
 \frac{2}{E v} \frac{q \mu}{b^{2}}+\frac{\pi}{2E v} \left(3+\frac{2}{v^{2}}\right)\frac{M q \mu}{b^{3}}\nn\\
 &+ \frac{2}{5E v}\left(24+\frac{50}{v^{2}}+\frac{5}{v^{4}}\right) \frac{M^{2} q \mu}{b^{4}}\nn\\
 &+\frac{3\pi}{2E^{2}v^2} \frac{q^{2} \mu^{2}}{b^{4}}+\mathcal{O}\left(\frac{[M]^5}{b^5}\right).
\end{align}
Finally, combining the above $\delta^K$ and $\delta^{\kappa_g}$, the total fourth order deflection angle can be written into a series form
\begin{align}
\label{deflection-angle}
\delta=\sum_{i=1}^{4}\delta^{(i)}+\mathcal{O}\left([M]^5/b^5\right),
\end{align}
with
\begin{align}
 \delta^{(1)}=&\frac{2M}{b}\left(1+\frac{1}{v^{2}}\right),\\
 \delta^{(2)}=&\frac{3 \pi}{4}\left(1+\frac{4}{v^{2}}\right) \frac{M^{2}}{b^{2}}+ \frac{2  sq\mu}{Eb^{2}v},\label{eq:dorder2}\\
 \delta^{(3)}=&\frac{2}{3}\left(5+\frac{45}{v^{2}}+\frac{15}{v^{4}}-\frac{1}{v^{6}}\right) \frac{M^{3}}{b^{3}}\nn\\
 &+\frac{\pi}{2 v}\left(5+\frac{4}{v^{2}}\right) \frac{  sq\mu M}{Eb^{3}},\\
 \delta^{(4)}=&\frac{105 \pi}{4}\left(\frac{1}{16}+\frac{1}{v^{2}}+\frac{1}{v^{4}}\right) \frac{M^{4}}{b^{4}}\nn\\
 &+
 \frac{6}{5 v}\left(18+\frac{45}{v^{2}}+\frac{5}{v^{4}}\right) \frac{  sq\mu M^{2}}{Eb^{4}}\nn\\
 &+\frac{3\pi}{2v^{2}} \frac{q^2\mu^2}{E^2b^{4}},
\end{align}
where the sign $s$ for the rotation direction has been restored. 

\section{Discussion of results}\label{Astrophysical-app}

\subsection{Effect of $\mu$ on deflection}
Setting $q=0$ in result \eqref{deflection-angle}, it reduces to the deflection of neutral massive particles in Schwarzschild spacetime (with or without electromagnetic fields)
\begin{align}
\label{Schwarzschild}
\delta_{S}=\sum_{i=1}^{4}\delta_{S}^{(i)}+\mathcal{O}\left(\frac{[M]^5}{b^5}\right),
\end{align}
with
\begin{align}
 \delta_{S}^{(1)}=&\frac{2M}{b}\left(1+\frac{1}{v^{2}}\right),\nn\\
 \delta_{S}^{(2)}=&\frac{3 \pi}{4}\left(1+\frac{4}{v^{2}}\right) \frac{M^{2}}{b^{2}},\nn\\
 \delta_{S}^{(3)}=&\frac{2}{3}\left(5+\frac{45}{v^{2}}+\frac{15}{v^{4}}-\frac{1}{v^{6}}\right) \frac{M^{3}}{b^{3}},\nn\\
 \delta_{S}^{(4)}=&\frac{105 \pi}{4}\left(\frac{1}{16}+\frac{1}{v^{2}}+\frac{1}{v^{4}}\right) \frac{M^{4}}{b^{4}},\nn
\end{align}
which agrees with the results in Refs.~\cite{LZLH2019,JiaKerr}. 

The deviation of the deflection angle with nonzero $\mu$ from pure Schwarzschild spacetime is therefore
\begin{align}
\delta_\mu=\delta-\delta_{S}=\sum_{i=1}^{4}\delta_\mu^{(i)}+\mathcal{O}\left(\frac{[M]^5}{b^5}\right),
\end{align}
with
\begin{align}
\delta_\mu^{(1)}=&0,\nn\\
 \delta_\mu^{(2)}=&\frac{2  sq\mu}{Eb^{2}v},\nn\\
\delta_\mu^{(3)}=&\frac{\pi}{2 v}\left(5+\frac{4}{v^{2}}\right) \frac{  sq\mu M}{Eb^{3}},\nn\\
 \delta_\mu^{(4)}=&\frac{6}{5 v}\left(18+\frac{45}{v^{2}}+\frac{5}{v^{4}}\right) \frac{  sq\mu M^{2}}{Eb^{4}}+\frac{3\pi}{2v^{2}} \frac{q^2\mu^2}{E^2b^{4}}.\nn
\end{align}
It is seen that the magnetic effect on the deflection appears from the second order (order $1/b^2$) of the impact parameter. Comparing to the effect of pure electric field on the deflection (see Eq. (4.4) of Ref. \cite{Xu:2021rld})
\begin{align}
    \delta_E=-\frac{2q}{Ev^2}\frac{Q}{M}\frac{M}{b}
\end{align}
where $Q$ is the total charge of the spacetime, we see that the $\delta_\mu$ is one order lower than $\delta_E$. 
This order comparison is also consistent with the effect of charge monopole and magnetic dipole on the deflection of charges in flat spacetime \cite{rutherfold}. 

On the other hand, the deflection by a central magnetic dipole is similar to the deflection by a rotating mass (i.e. the Kerr spacetime) in at least the following ways. Firstly, comparing to Schwarzschild spacetime, both scenarios assert an extra axisymmetric force field on the charged particle. Secondly, the $g_{rr}$ component of the Kerr metric with angular momentum per unit mass $a$ contains a term asymptotically proportional to $-a^2/r^2\sim \Phi_0(r)$ yielding a force $ |\nabla \Phi_0(r)|\sim a^2/r^3$. While the magnetic dipole \eqref{eq:flatmag} generates asymptotically the Lorentz force $\sim qv|\mathbf{B}|\sim q\mu/r^3$. Therefore to the leading order, we should expect that the effect of the magnetic dipole on the radial motion of charged signals should resemble that of the spacetime spin to a neutral signal.  
Indeed, the deflection of neutral signal in Kerr spacetime has been known to order four too ~\cite{JiaKerr}. To the second order, i.e., the order $a$ first appears, this deflection is
\begin{align}
\delta_{K}=&\frac{2M}{b}\left(1+\frac{1}{v^{2}}\right)+\frac{3 \pi}{4}\left(1+\frac{4}{v^{2}}\right) \frac{M^{2}}{b^{2}}\nn\\
&-\frac{4M (sa)}{b^2v}+\mathcal{O}\left( \frac{|M|^3}{b^3}\right).
\end{align}
Comparing to Eq. \eqref{eq:dorder2}, we see that from the deflection angle point of view, the equivalence relation between dipole and Kerr spacetime spin, to the leading order is
\begin{align}
    q\mu\to -2EM\cdot a. \label{eq:aqmurel}
\end{align}

\subsection{Gravitational lensing}

With the analogy \eqref{eq:aqmurel} between the Schwarzschild magnetic dipole result and Kerr spacetime result of the deflection $\delta$, it is natural to expect that the GL in the Schwarzschild magnetic dipole case is similar to the Kerr case: after all, the GL equation usually is only solved using the deflection angle to the first one or two orders. If to these orders, the GL equation and formula for images' apparent angles in these two cases are also the same, then we will be able to directly use the results obtained in Kerr spacetime \cite{Huang:2020trl, Zhang:2022tbp} for the images' apparent angles $\theta_{Sm}$ in the current case. Indeed this is the case for both the GL equation (see Eq. (37) of Ref. \cite{Li:2019qyb} and Eq. (64) and (B1) of Ref. \cite{Zhang:2022tbp}) and the formula for apparent angles (see Eq. (71) of Ref. \cite{Zhang:2022tbp}). 
Therefore the apparent angle for the images can be directly quoted from Eq. (82) of Ref. \cite{Zhang:2022tbp}
\begin{align}
    \theta_{Sm}=\frac{b_{0s}}{r_d}+\frac{b_{1s}}{r_d}
+\mathcal{O}\left( \frac{b_{0s}^3}{r_d^3}\right),
\label{eq:thetakerr}
\end{align}
where $b_{0s}$ and $b_{1s}$ are the leading and next-leading order impact parameters 
\begin{subequations}
\label{eq:kbkdefs}
\begin{align}
b_{0s}=&\frac{\varphi_0 r_d r_s}{2\lb r_d+r_s\rb}\lb \sqrt{ 1+\eta } -s \rb,\\
b_{1s}=&\frac{\eta \lsb 8 s q\mu v/(EM)+3M \pi \lb 4+v^2 \rb \rsb}{32\lb 1+v^2\rb\sqrt{
1+ \eta }\lb \sqrt{
1+ \eta }-s\rb} ,\label{eq:kb1}\\
\eta =&\frac{8M\lb r_d +r_s \rb}{\varphi^2_0 r_d r_s } \lb 1+ \frac{1}{v^2} \rb. \label{eq:kbkdefs3}
\end{align}
\end{subequations}

There are a few properties of these apparent angles worth mentioning. The first is that the effect of the charge-magnetic dipole interaction on the apparent angles is proportional to $q\mu/E$, but not directly on the kind of the charge as long as they are highly relativistic. For example, no matter whether they are electrons, protons, or even light nuclei in the cosmic rays, their effect only depends on their $q/E$ values once the particles are highly relativistic. The second point to note is that as in the deflection angle, the magnetic dipole influences the apparent angels of the charged signals from the next-leading order too. This also implies that because Eq. \eqref{eq:thetakerr} is a perturbative result, it is only valid when $b_{1s}\ll b_{0s}$. Using the large and small $\eta$ limits of Eq. \eqref{eq:kbkdefs}, this condition further restricts the applicable parameter space to the case
\begin{align}
    q\mu/E\ll \mathrm{min}\left\{ \sqrt{M^3 r_{s,d}},~M^2/\varphi_0,~r_{s,d}^2\phi_0^3. \right\} 
    \label{eq:qmuebd}
\end{align}
We see that the higher the energy of the charges, the more applicable the above results.

\begin{figure}
    \centering
    \includegraphics[width=0.4\textwidth]{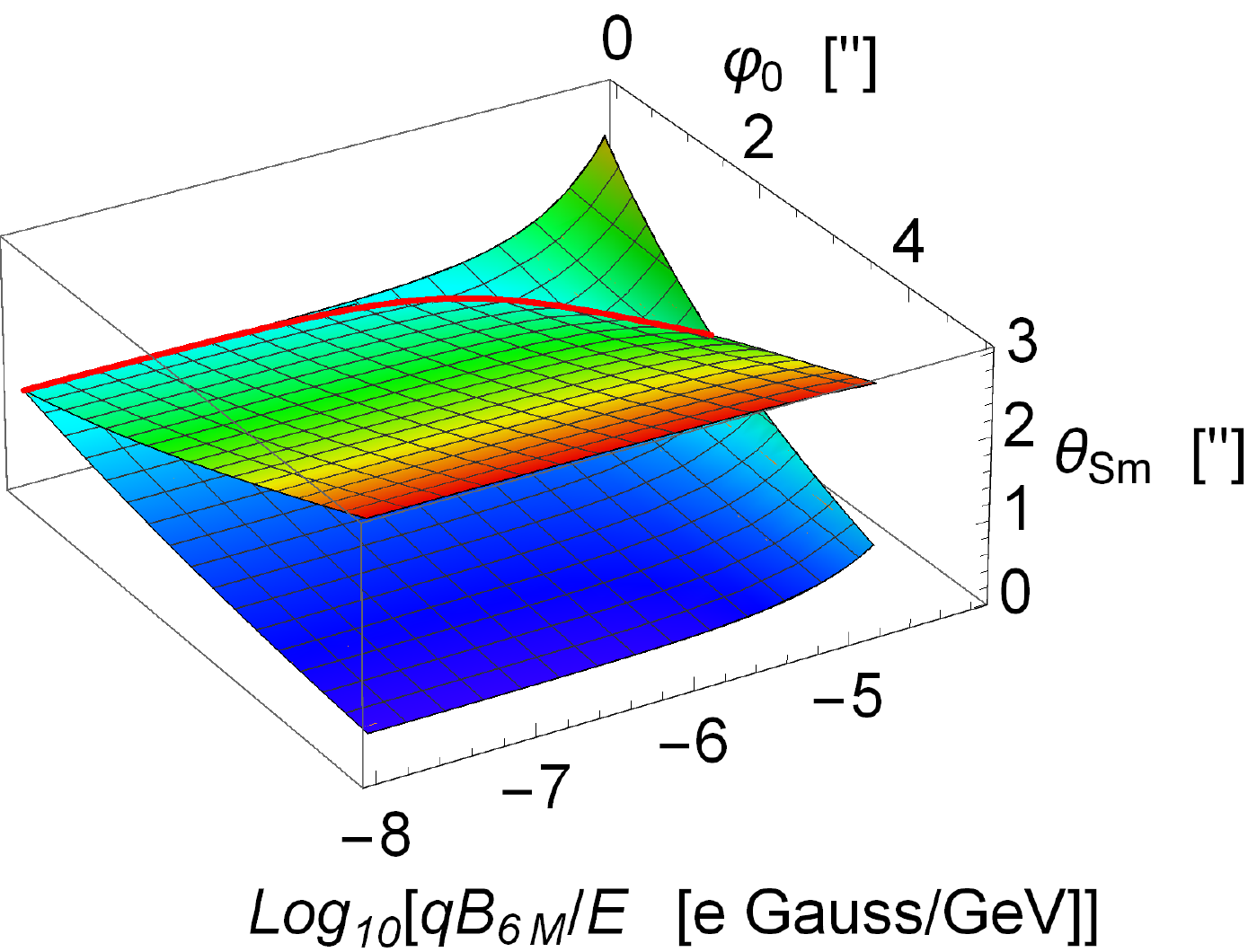}\\
    (a)\\
    \includegraphics[width=0.4\textwidth]{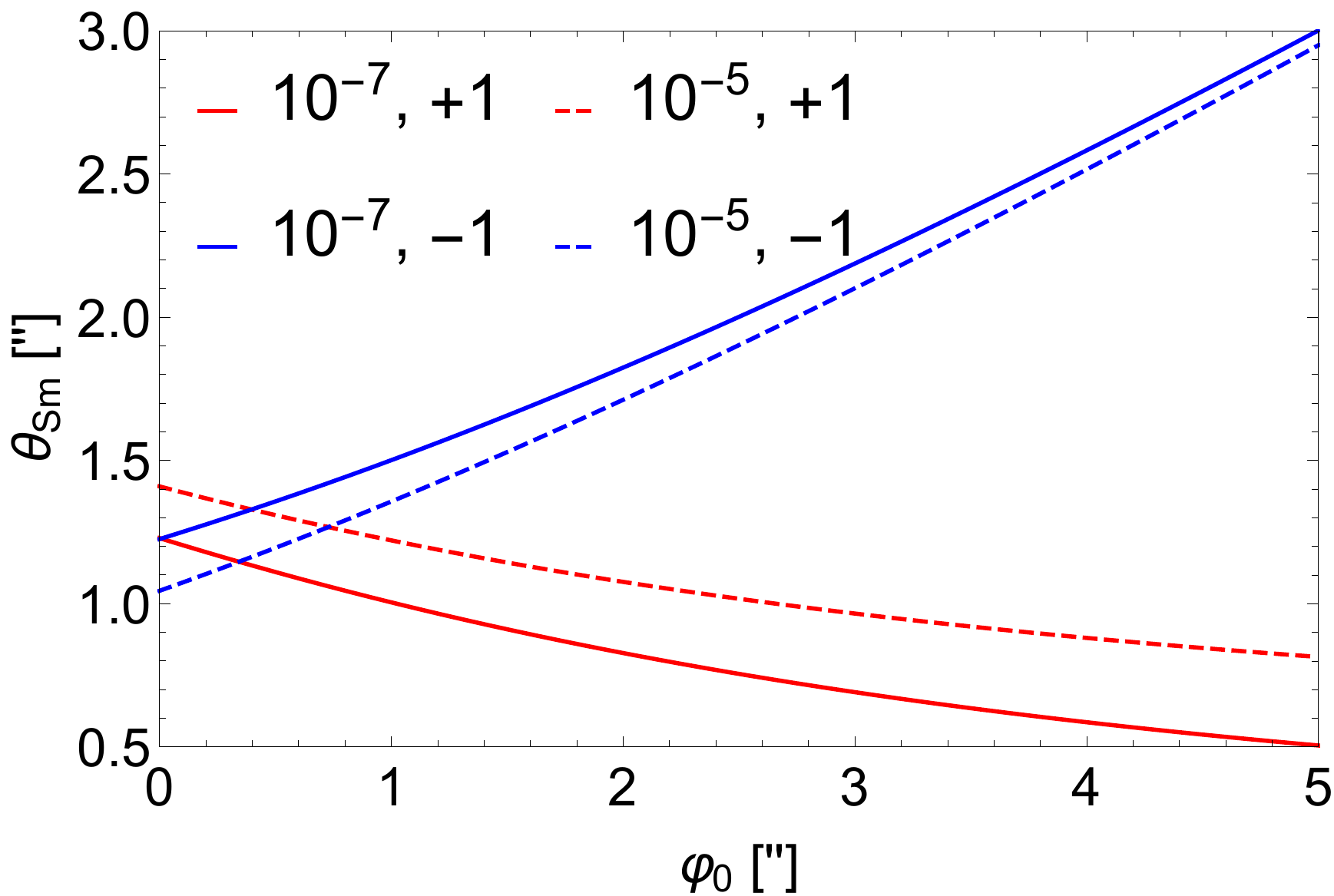}\\
    (b)\\
    \includegraphics[width=0.4\textwidth]{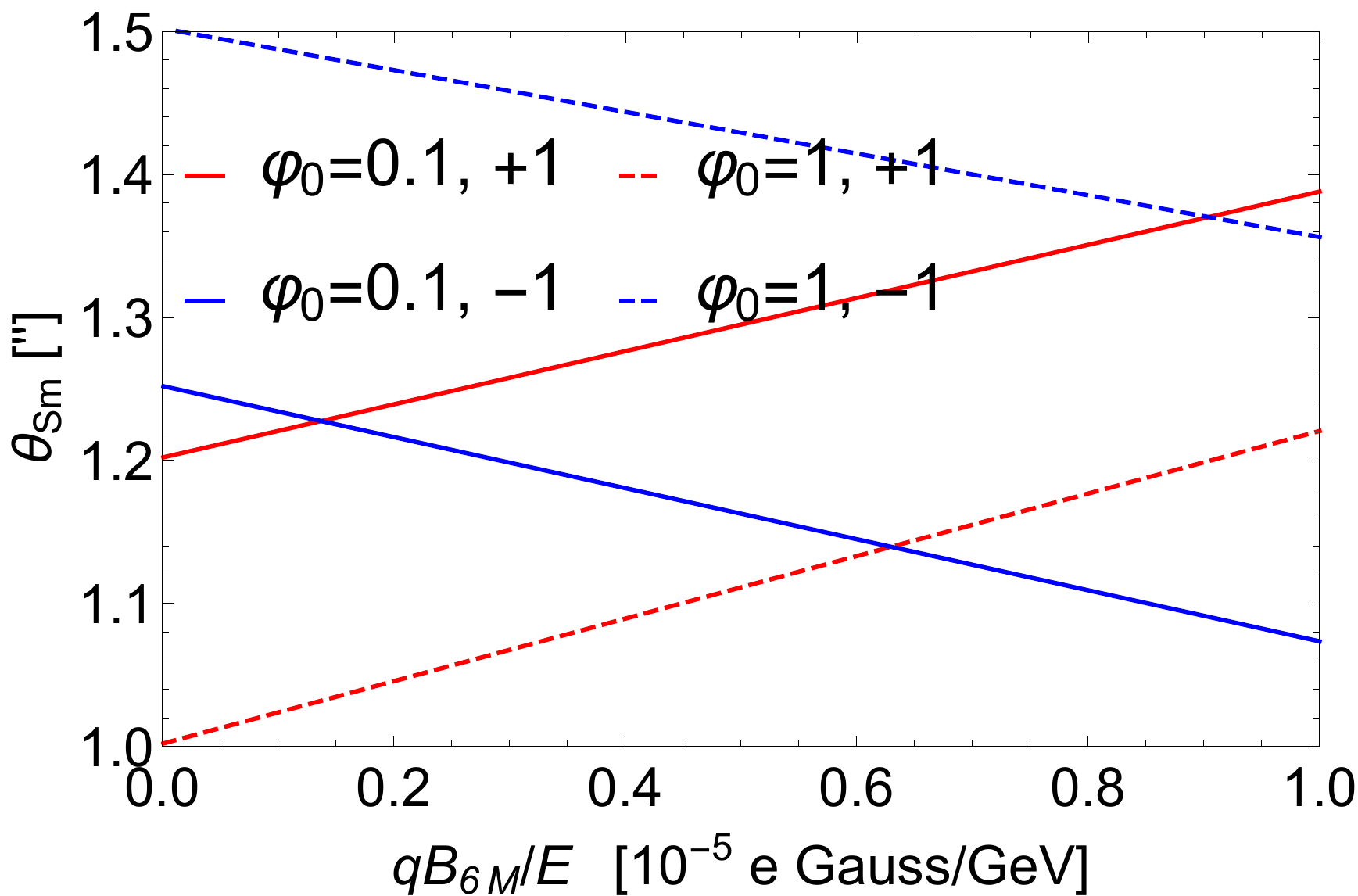}\\
    (c)
    \caption{The apparent angles $\theta_{Sm}$ using Eq. \eqref{eq:thetakerr} for M87$^*$. (a) 3D plot of $\theta_{Sm}$ as a function of $qB_{6M}/E$ and $\varphi_0$. Note there exists an upper boundary (red curve) for $q\mu/E$ determined by Eq. \eqref{eq:qmuebd}; (b) $\theta_{Sm}$ as a function of $\varphi_0$ for $qB_{6M}/E=10^{-7}$ [e Gauss/GeV] (solid curves) and $qB_{6M}/E=10^{-5}$ [e Gauss/GeV] (dash curves) for two directions $s=+1$ (red curves) and $s=-1$ (blue curves); (c) $\theta_{Sm}$ as a function of $qB_{6M}/E$ for $\varphi_0=0.1$ [$^{\prime\prime}$] (solid curves) and $\varphi_0=1$ [$^{\prime\prime}$] (dash curves) for  $s=+1$ (red curves) and $s=-1$ (blue curves).
    }
    \label{fig:lensaang}
\end{figure}

\begin{figure}
    \centering
    \includegraphics[width=0.4\textwidth]{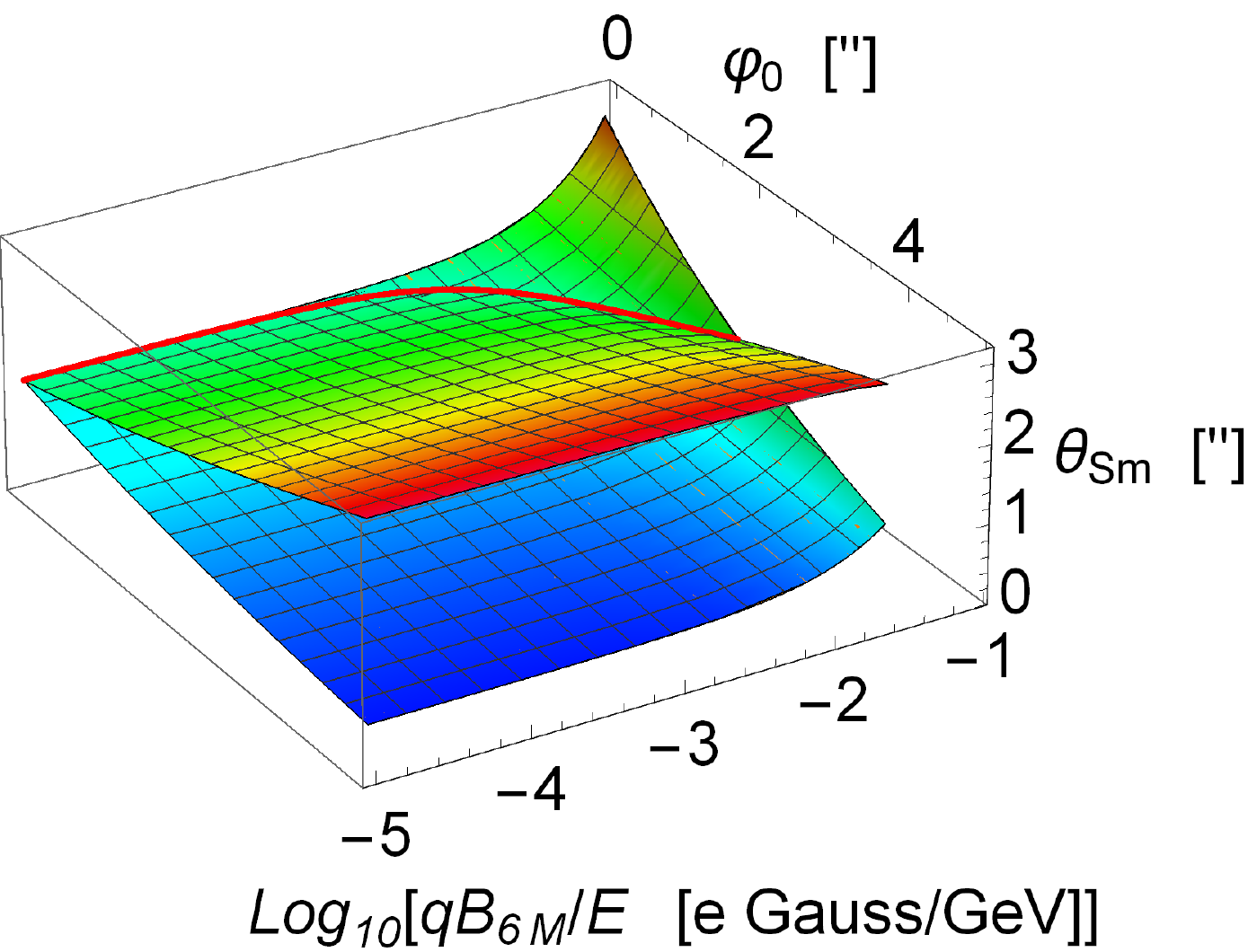}
    \\
    (a)\\
    \includegraphics[width=0.4\textwidth]{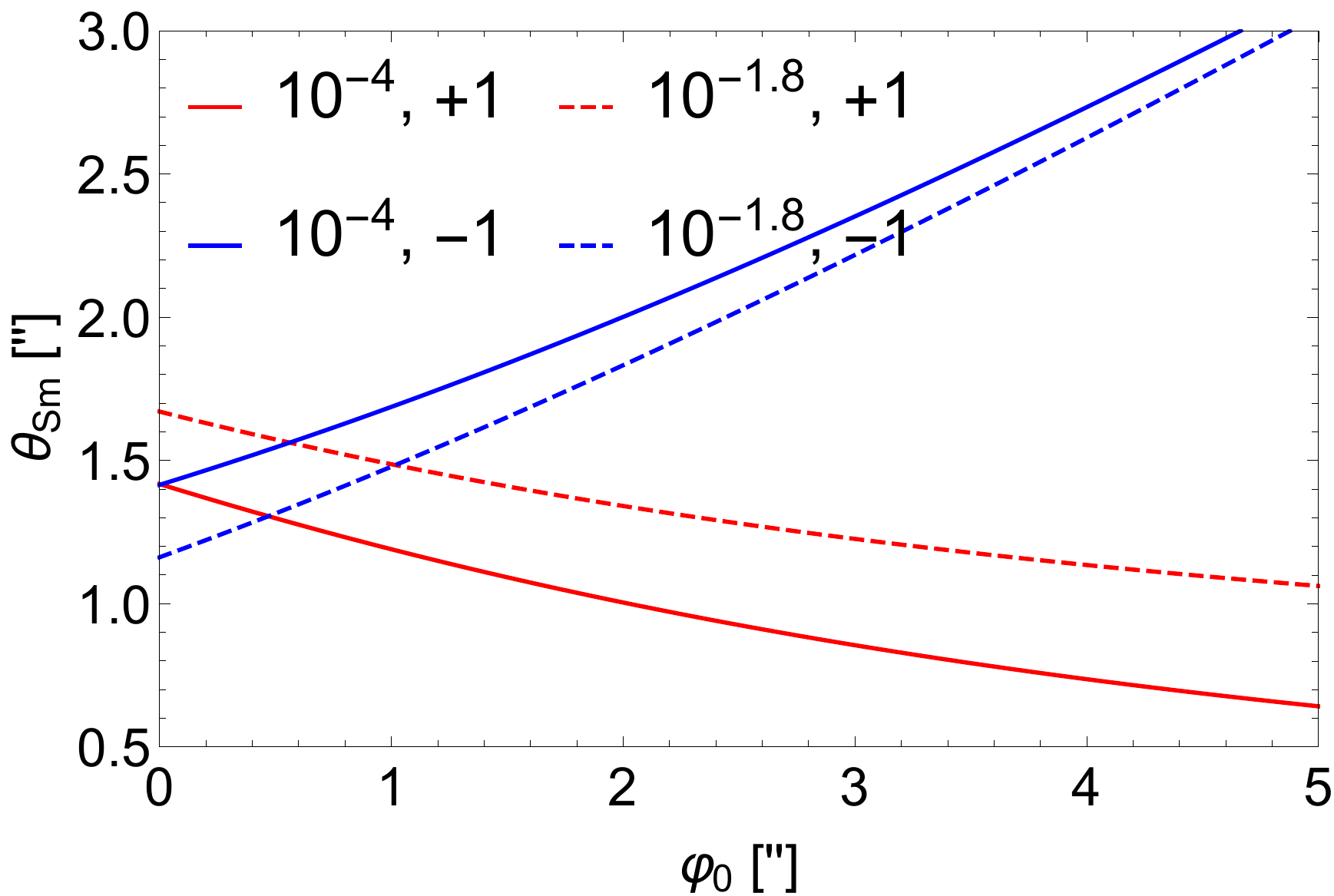}\\
    (b)\\
    \includegraphics[width=0.4\textwidth]{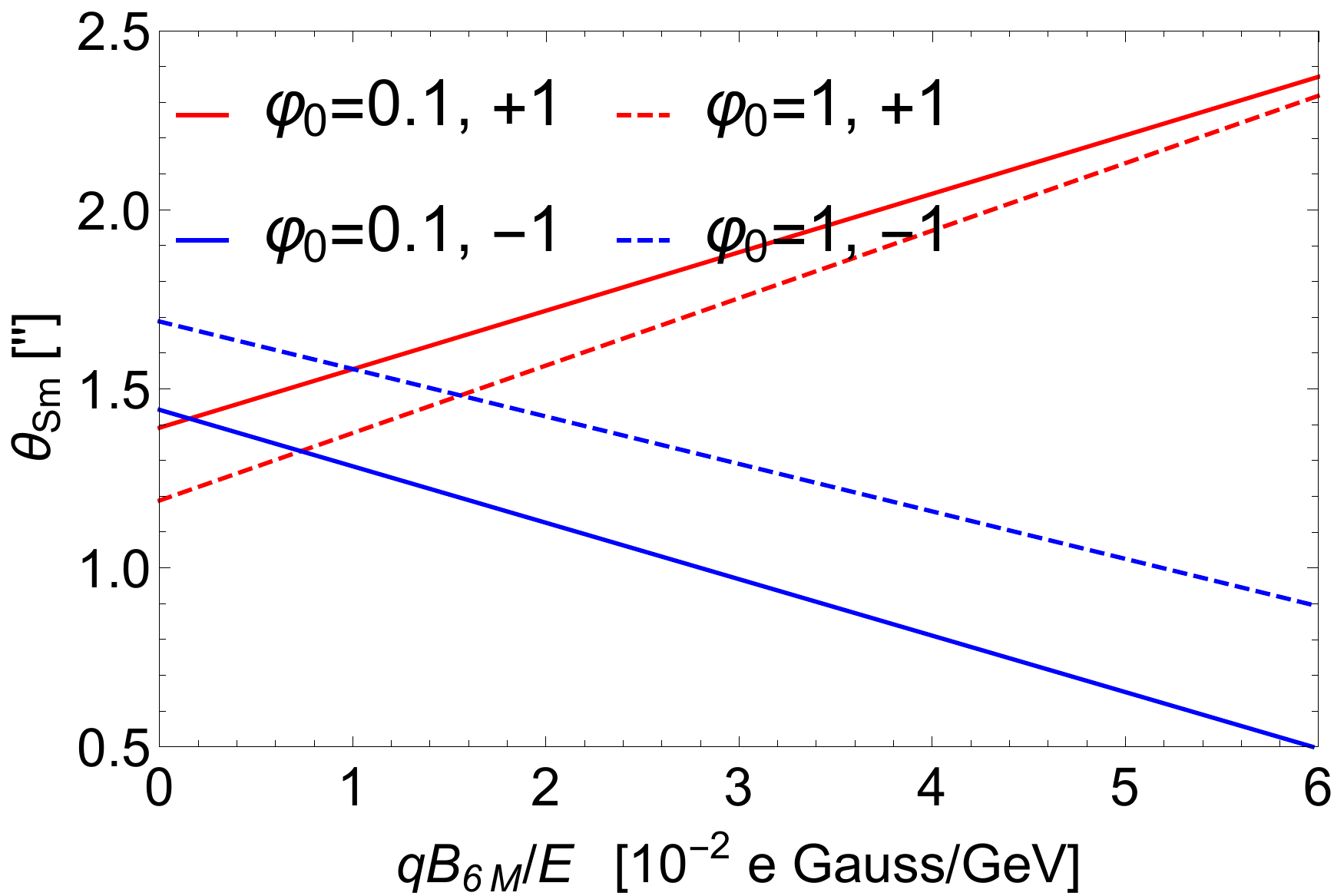}\\
    (c)
    \caption{The apparent angles $\theta_{Sm}$ using Eq. \eqref{eq:thetakerr} for Sgr A$^*$. (a) 3D plot of $\theta_{Sm}$ as a function of $qB_{6M}/E$ and $\varphi_0$. Note there exists an upper boundary (red curve) for $q\mu/E$ determined by Eq. \eqref{eq:qmuebd}; (b) $\theta_{Sm}$ as a function of $\varphi_0$ for $qB_{6M}/E=10^{-4}$ [e Gauss/GeV] (solid curves) and $qB_{6M}/E=10^{-1.8}$ [e Gauss/GeV] (dash curves) for  $s=+1$ (red curves) and $s=-1$ (blue curves); (c) $\theta_{Sm}$ as a function of $qB_{6M}/E$ for $\varphi_0=0.1$ [$^{\prime\prime}$] (solid curves) and $\varphi_0=1$ [$^{\prime\prime}$] (dash curves) for  $s=+1$ (red curves) and $s=-1$ (blue curves).}
    \label{fig:lensaang2}
\end{figure}

To obtain a better intuition of the effect of $\mu$ on the observations, in Fig. \ref{fig:lensaang} and \ref{fig:lensaang2} we plot the apparent angles \eqref{eq:thetakerr} as a function of $q\mu/E$ and $\phi_0$. For the charge $q$ of cosmic ray particle, it can only choose a few discrete values from $1e$ of protons/deuterons/tritons to $\sim 26 e$ for iron nuclei. For the energy $E$, we concentrate on the range $E\gtrsim 10^3$ [GeV]. 
We will choose the Sgr A$^*$ and M87$^*$ supermassive black holes as the lens, and assume the source of the signal roughly is located at the same radius as the detector. 
For the magnetic dipole moment, we assume that the current is due to the accretion materials near the innermost stable circular orbit with $R_0=6M$. For M87$^*$, there is already a rough estimate of 1 [Gauss] to 30 [Gauss] for the magnetic field $B_{6M}$ at this radius \cite{EventHorizonTelescope:2021srq}, while for Sgr A$^*$, this magnetic field is only known to be tens of Gauss \cite{Johnson:2015iwg}. Using \eqref{eq:btheta} with this magnetic field, we will be able to deduce the corresponding $\mu$ as
\begin{align}
    \mu=\frac{8\sqrt{6}M^3B_{6M}}{5-12\ln(3/2)}. \label{eq:muinb}
\end{align}
That is, $\mu$ will be strictly proportional to $B_{6M}$.

Fig. \ref{fig:lensaang} plots the $\theta_s$ for M87$^*$ SMBH as functions of $\phi_0$ and $qB_{6M}/E$.
From Fig. \ref{fig:lensaang} (a) it is seen that when $qB_{6M}/E$ is smaller than a rough value $\lambda_{M87}\approx 10^{-6}$ [e Gauss/GeV], its effect on the the apparent angles of the two images $s=\pm1$ are negligible. In other words, the magnetic interaction is still weaker than gravitational deflection. When $qB_{6M}/E$ exceeds this value however, the magnetic interaction can grow stronger than the gravitational field, and consequently the apparent angle from the counterclockwise direction is decreased dramatically by the magnetic dipole while that from the opposite direction is increased, as can be more clearly seen from Fig. \ref{fig:lensaang} (b). This dependence is also consistent with the effect of $\mu$ on the deflection angle. 
This value $\lambda_{M87}$ ($10^{-6}$ [e Gauss/GeV]) for $qB_{6M}/E$ is expected to be reachable by many cosmic rays since there are plenty of signals above the so-called ``knee'' structure around $10^{-6.6}$ [GeV] in the cosmic ray spectrum  \cite{Anchordoqui:2018qom} and the magnetic field $B_{6M}$ is expected to in the order of tens of Gauss. Therefore measuring the dependence of the apparent angles of such charged signals near the M87 galaxy center will help to constrain the exact value of the magnetic field. Indeed, using even higher energy cosmic rays such as those above the ``second knee'' ($10^{8.0}$ [GeV]) or the ``ankle'' ($10^{9.7}$ [GeV]), the magnetic field as weak as $10^{-4}$ [Gauss] near the accretion radius can also be constrained. 

Fig. \ref{fig:lensaang} (c) shows the apparent angles for two typical source angular positions $\phi_0$ using $qB_{6M}/E$ as the $x$-axis. The variation of $\theta_{Sm}$ in this plot means that for each fixed kind of charged signal, as their energy $E$ varies, the apparent angles from each side of the lens will also vary. That is, non-mono-energetic charged signals will form extended images, as long as $B_{6M}$ is not too small. It is seen that all the four curves are strictly linear to $qB_{6M}/E$, as dictated by Eqs. \eqref{eq:kb1} and \eqref{eq:muinb}. Indeed, from these equations we can find the slope of these curves, in the relativistic and $\phi_0$ (i.e., small $\phi_0$) limits, to be
\begin{align}
    k=\frac{\sqrt{6}sM^2}{\lsb 5-12\ln(3/2)\rsb r_d} .
\end{align}
This is also consistent with the observation in this plot that basically all slopes are of the same absolute size, which are determined by $M$ and $r_d$ of the system, but not by $\phi_0$.

Fig. \ref{fig:lensaang2} illustrates the apparent angles for the Sgr A$^*$ SMBH. It is seen that qualitatively they are similar to the M87$^*$ case, except the critical value of $q\mu/E$ in this case is now about $\lambda_{SgrA}\approx 10^{-3}$ [e Gauss/GeV]. This implies that cosmic rays with the same energy are more easily affected by the magnetic dipole of the same strength around the Sgr A$^*$ SMBH. For the expected $\mathcal{O}(10)$ [Gauss] magnetic field around the accretion radius, the cosmic ray with energy as low as $10^3$ [GeV] will be able to experience the large effect of the magnetic field on its apparent angles. In other words, the cosmic rays above the ``knee'' will be able to detect the magnetic field as low as $10^{-3}$ [Gauss] at the accretion radius of Sgr A$^*$. The other qualitative features in the two-dimensional plots in Fig. \ref{fig:lensaang2} (b) and (c) are the same as in Fig. \ref{fig:lensaang}.

\section{Concluding Remarks}\label{conclusion}

In this paper, we have studied the deflection angle of a charged particle in the equatorial plane of a Schwarzschild spacetime with a dipole magnetic field, using the GB theorem with the generalized Jacobi metric method and the osculating Riemannian metric method. The fact that the deflection angles for trajectories in the clockwise and anticlockwise directions are different manifests the non-reversibility of the Finsler metric. It is found that the magnetic dipole $\mu$ will decrease (or increase) the deflection angle of a positively charged signal if its rotation angular momentum is parallel (or antiparallel) to the magnetic field. To the leading order it appears, the effect of $\mu$ on the deflection is similar to the effect of spacetime spin in Kerr spacetime. This similarity allows us to solve the GL equation and study the effect of $\mu$ on the apparent angles of the GL images. 
Applications of these results to M87$^*$ and Sgr A$^*$ SMBHs suggest that by measuring the apparent angles of high-energy cosmic rays, the magnetic field around these SMBHs might be constrained. 

We point out that the above conclusion does not have to rely on the assumption that the magnetic dipole is generated by the accretion materials near the innermost stable circular orbit, although we used this as an example. Indeed, as long as there exists a magnetic dipole, an apparent angle figure similar to Fig. \ref{fig:lensaang} (a) and \ref{fig:lensaang2} (a) can always be drawn (replacing $qB_{6M}/E$ by $q\mu/E$), and therefore the dipole can still be constrained by the observation of the apparent angles. 

Regarding future direction, three questions are particularly interesting. The first is whether the analogy between the magnetic dipole and Kerr spacetime spin can be extended to other quantities about the particle's motion, such as its total travel time and the time delay between images in GL. To answer this question, a perturbative computation seems unavoidable since the geometric method is only applicable to the calculation of the deflection angle but not the travel time. 
The second is to generalize the method in this work to the Kerr case because both the black hole's spin and dipole magnetic field have intricate effects on charged particles and their addition or competition might be interesting. 
The third and more challenging direction is to study 
the coupling between the particle spin and the magnetic field in a curve background since these two are well known to be coupled even in flat spacetime. It is expected that their coupling in a curved spacetime might bring more complex and interesting features in the particle's motion, and deflection in particular.

\acknowledgements

The authors thank Mr. Jihong He and Tingyuan Jiang for the illustration of some figures. This work is supported by the National Key Research and Development Program of China (Grants No. 2021YFA0718500, 2021YFA0718503) and the NSFC (12133007, U1838103).

\end{document}